\input harvmac
\overfullrule=0pt
\parindent 25pt
\tolerance=10000

\input epsf

\newcount\figno
\figno=0
\def\fig#1#2#3{
\par\begingroup\parindent=0pt\leftskip=1cm\rightskip=1cm\parindent=0pt
\baselineskip=11pt
\global\advance\figno by 1
\midinsert
\epsfxsize=#3
\centerline{\epsfbox{#2}}
\vskip 12pt
{\bf Fig.\ \the\figno: } #1\par
\endinsert\endgroup\par
}
\def\figlabel#1{\xdef#1{\the\figno}}
\def\encadremath#1{\vbox{\hrule\hbox{\vrule\kern8pt\vbox{\kern8pt
\hbox{$\displaystyle #1$}\kern8pt}
\kern8pt\vrule}\hrule}}

 \def\frac#1#2{{#1\over #2}}

 \def\s{\sqrt}

 \def\CO{{\cal O}}

 \def\al{\alpha'}
 \def\de{\partial}

 \def\f {\frac}
 \def\ti{\tilde}
 \def\ap{\alpha}

 \def\ddd{\cdot\cdot\cdot}
 
 \def\la{\langle}
 \def\lb{\rangle}

 \def\frac#1#2{{#1\over #2}}

 \def\s{\sqrt}

 \def\CO{{\cal O}}

 \def\nh{\textstyle \frac{|n|}{2}}
 
 \def\ha{\textstyle \frac{1}{2}}
 \def\nhk{\textstyle \frac{k n'}{2}}

\font\blackboard=msbm10 \font\blackboards=msbm7
\font\blackboardss=msbm5
\newfam\black
\textfont\black=\blackboard
\scriptfont\black=\blackboards
\scriptscriptfont\black=\blackboardss
\def\blackb#1{{\fam\black\relax#1}}


\lref\KKK{
V.~Kazakov, I.~K.~Kostov and D.~Kutasov,
``A matrix model for the two-dimensional black hole,''
Nucl.\ Phys.\ B {\bf 622}, 141 (2002)
[arXiv:hep-th/0101011].
}

\lref\DVV{
R.~Dijkgraaf, H.~Verlinde and E.~Verlinde,
``String propagation in a black hole geometry,''
Nucl.\ Phys.\ B {\bf 371}, 269 (1992).
}

\lref\ST{
A.~Strominger and T.~Takayanagi,
``Correlators in timelike bulk Liouville theory,''
Adv.\ Theor.\ Math.\ Phys.\  {\bf 7}, 369 (2003)
[arXiv:hep-th/0303221].
}

\lref\SO{
A.~Strominger,
``Open string creation by S-branes,''
arXiv:hep-th/0209090.
}

\lref\GS{
M.~Gutperle and A.~Strominger,
``Timelike boundary Liouville theory,''
Phys.\ Rev.\ D {\bf 67}, 126002 (2003)
[arXiv:hep-th/0301038].
}

\lref\Ma{
B.~C.~Da Cunha and E.~J.~Martinec,
``Closed string tachyon condensation and worldsheet inflation,''
Phys.\ Rev.\ D {\bf 68}, 063502 (2003)
[arXiv:hep-th/0303087].
}

\lref\SC{
V.~Schomerus,
``Rolling tachyons from Liouville theory,''
JHEP {\bf 0311}, 043 (2003)
[arXiv:hep-th/0306026].
}

\lref\FH{
T.~Fukuda and K.~Hosomichi,
``Three-point functions in sine-Liouville theory,''
JHEP {\bf 0109}, 003 (2001)
[arXiv:hep-th/0105217].
}

\lref\HK{
K.~Hori and A.~Kapustin,
``Duality of the fermionic 2d black hole and $N = 2$
Liouville theory as  mirror symmetry,''
JHEP {\bf 0108}, 045 (2001)
[arXiv:hep-th/0104202].
}

\lref\CKR{
B.~Craps, D.~Kutasov and G.~Rajesh,
``String propagation in the presence of cosmological singularities,''
JHEP {\bf 0206}, 053 (2002)
[arXiv:hep-th/0205101].
}

\lref\Marcus{
N.~Marcus,
``Unitarity and regularized divergences in string amplitudes,''
Phys.\ Lett.\ B {\bf 219}, 265 (1989).
}

\lref\WW{
E.~J.~Weinberg and A.~q.~Wu,
``Understanding complex perturbative effective potentials,''
Phys.\ Rev.\ D {\bf 36}, 2474 (1987).
}

\lref\CKL{
B.~Craps, P.~Kraus and F.~Larsen,
``Loop corrected tachyon condensation,''
JHEP {\bf 0106}, 062 (2001)
[arXiv:hep-th/0105227].
}

\lref\HPT{
A.~Hanany, N.~Prezas and J.~Troost,
``The partition function of the two-dimensional black hole conformal  field
theory,''
JHEP {\bf 0204}, 014 (2002)
[arXiv:hep-th/0202129].
}

\lref\MOS{
J.~M.~Maldacena, H.~Ooguri and J.~Son,
``Strings in $AdS_3$ and the $SL(2,\blackb{R})$ WZW model. II:
  Euclidean black hole,''
J.\ Math.\ Phys.\  {\bf 42}, 2961 (2001)
[arXiv:hep-th/0005183].
}

\lref\BLW{
A.~Buchel, P.~Langfelder and J.~Walcher,
``On time-dependent backgrounds in supergravity and string theory,''
Phys.\ Rev.\ D {\bf 67}, 024011 (2003)
[arXiv:hep-th/0207214].
}

\lref\Wi{
E.~Witten,
``On string theory and black holes,''
Phys.\ Rev.\ D {\bf 44}, 314 (1991).
}

\lref\MSW{
G.~Mandal, A.~M.~Sengupta and S.~R.~Wadia,
``Classical solutions of two-dimensional string theory,''
Mod.\ Phys.\ Lett.\ A {\bf 6}, 1685 (1991).
}

\lref\Pi{
M.~J.~Perry and E.~Teo,
``Nonsingularity of the exact two-dimensional string black hole,''
Phys.\ Rev.\ Lett.\  {\bf 70}, 2669 (1993)
[arXiv:hep-th/9302037];
P.~Yi,
``Nonsingular 2-D black holes and classical string backgrounds,''
Phys.\ Rev.\ D {\bf 48}, 2777 (1993)
[arXiv:hep-th/9302070].
}

\lref\TSQ{
A.~A.~Tseytlin,
``Effective action of gauged WZW model and exact string solutions,''
Nucl.\ Phys.\ B {\bf 399}, 601 (1993)
[arXiv:hep-th/9301015].
}

\lref\PS{
T.~Suyama and P.~Yi,
``A holographic view on matrix model of black hole,''
JHEP {\bf 0402}, 017 (2004)
[arXiv:hep-th/0401078].
}

\lref\TV{
A.~A.~Tseytlin and C.~Vafa,
``Elements of string cosmology,''
Nucl.\ Phys.\ B {\bf 372}, 443 (1992)
[arXiv:hep-th/9109048].
}

\lref\LU{
D.~Lust,
``Cosmological string backgrounds,''
arXiv:hep-th/9303175.
}

\lref\KL{
C.~Kounnas and D.~Lust,
``Cosmological string backgrounds from gauged WZW models,''
Phys.\ Lett.\ B {\bf 289}, 56 (1992)
[arXiv:hep-th/9205046].
}

\lref\GKK{
A.~Giveon and D.~Kutasov,
``Notes on $AdS_3$,''
Nucl.\ Phys.\ B {\bf 621}, 303 (2002)
[arXiv:hep-th/0106004].
}

\lref\GK{
A.~Giveon and D.~Kutasov,
``Comments on double scaled little string theory,''
JHEP {\bf 0001}, 023 (2000)
[arXiv:hep-th/9911039].
}

\lref\APS{
A.~Adams, J.~Polchinski and E.~Silverstein,
``Don't panic! Closed string tachyons in ALE space-times,''
JHEP {\bf 0110}, 029 (2001)
[arXiv:hep-th/0108075].
}

\lref\Sen{
A.~Sen,
``Rolling tachyon,''
JHEP {\bf 0204}, 048 (2002)
[arXiv:hep-th/0203211];
``Tachyon matter,''
JHEP {\bf 0207}, 065 (2002)
[arXiv:hep-th/0203265].
}

\lref\LNT{
F.~Larsen, A.~Naqvi and S.~Terashima,
``Rolling tachyons and decaying branes,''
JHEP {\bf 0302}, 039 (2003)
[arXiv:hep-th/0212248].
}

\lref\KSMC{
J.~L.~Karczmarek and A.~Strominger,
``Matrix cosmology,''
JHEP {\bf 0404}, 055 (2004)
[arXiv:hep-th/0309138].
}

\lref\KSMCC{
J.~L.~Karczmarek and A.~Strominger,
``Closed string tachyon condensation at $c = 1$,''
JHEP {\bf 0405}, 062 (2004)
[arXiv:hep-th/0403169];

S.~R.~Das, J.~L.~Davis, F.~Larsen and P.~Mukhopadhyay,
``Particle production in matrix cosmology,''
arXiv:hep-th/0403275;

J.~L.~Karczmarek, A.~Maloney and A.~Strominger,
``Hartle-Hawking vacuum for c = 1 tachyon condensation,''
arXiv:hep-th/0405092.
}

\lref\LLM{
N.~Lambert, H.~Liu and J.~Maldacena,
``Closed strings from decaying D-branes,''
arXiv:hep-th/0303139.
}

\lref\KLMS{
J.~L.~Karczmarek, H.~Liu, J.~Maldacena and A.~Strominger,
``UV finite brane decay,''
JHEP {\bf 0311}, 042 (2003)
[arXiv:hep-th/0306132].
}

\lref\CO{
L.~Cornalba and M.~S.~Costa,
``Time-dependent orbifolds and string cosmology,''
Fortsch.\ Phys.\  {\bf 52}, 145 (2004)
[arXiv:hep-th/0310099].
}

\lref\AA{
J.~Teschner,
``On structure constants and fusion rules in the $SL(2,\blackb{C})/SU(2)$
  WZNW  model,''
Nucl.\ Phys.\ B {\bf 546}, 390 (1999)
[arXiv:hep-th/9712256].
}

\lref\AB{
J.~Teschner,
``Operator product expansion and factorization in the $H_3^+$ WZNW model,''
Nucl.\ Phys.\ B {\bf 571}, 555 (2000)
[arXiv:hep-th/9906215].
}

\lref\mini{
S.~Fredenhagen and V.~Schomerus,
``On minisuperspace models of S-branes,''
JHEP {\bf 0312}, 003 (2003)
[arXiv:hep-th/0308205].
}

\lref\minitwo{
Y.~Hikida,
``String theory on Lorentzian $AdS_3$ in minisuperspace,''
JHEP {\bf 0404}, 025 (2004)
[arXiv:hep-th/0403081].
}

\lref\HMT{
E.~J.~Martinec,
``Defects, decay, and dissipated states,''
arXiv:hep-th/0210231;
M.~Headrick, S.~Minwalla and T.~Takayanagi,
``Closed string tachyon condensation: An overview,''
Class.\ Quant.\ Grav.\  {\bf 21}, S1539 (2004)
[arXiv:hep-th/0405064].
}

\lref\Gaw{
K.~Gawedzki,
``Noncompact WZW conformal field theories,''
arXiv:hep-th/9110076.
}

\lref\KuHu{
D.~Kutasov,
``Irreversibility of the renormalization
group flow in two-dimensional quantum gravity,''
Mod.\ Phys.\ Lett.\ A {\bf 7}, 2943 (1992) [arXiv:hep-th/9207064];
E.~Hsu and D.~Kutasov,
``The gravitational sine-Gordon model,''
Nucl.\ Phys.\ B {\bf 396}, 693 (1993) [arXiv:hep-th/9212023].
}

\lref\TSM{
J.~Teschner,
``The mini-superspace limit of the $SL(2,\blackb{C})/SU(2)$ WZNW model,''
Nucl.\ Phys.\ B {\bf 546}, 369 (1999)
[arXiv:hep-th/9712258].
}

\lref\SHM{
S.~Ribault and V.~Schomerus,
``Branes in the 2-D black hole,''
JHEP {\bf 0402}, 019 (2004) [arXiv:hep-th/0310024].
}

\lref\ZZL{
A.~B.~Zamolodchikov and A.~B.~Zamolodchikov,
``Structure constants and conformal bootstrap in Liouville field theory,''
Nucl.\ Phys.\ B {\bf 477}, 577 (1996) [arXiv:hep-th/9506136].
}

\lref\Seiberg{
N.~Seiberg,
``Notes On Quantum Liouville Theory And Quantum Gravity,''
Prog.\ Theor.\ Phys.\ Suppl.\  {\bf 102}, 319 (1990).
}

\lref\KlR{
I.~R.~Klebanov,
``String theory in two-dimensions,''
arXiv:hep-th/9108019.
}

\lref\Is{
N.~Ishibashi, K.~Okuyama and Y.~Satoh,
``Path integral approach to string theory on $AdS_3$,''
Nucl.\ Phys.\ B {\bf 588}, 149 (2000)
[arXiv:hep-th/0005152].
}

\lref\TTL{T.~Takayanagi, ``Matrix model and time-like linear
dilaton matter,'' arXiv:hep-th/0411019.
}

\baselineskip 18pt plus 2pt minus 2pt

\Title{\vbox{\baselineskip12pt
\hbox{hep-th/0408124}
\hbox{SNUST-040801}
  }}
{\vbox{\centerline{On Solvable Time-Dependent Model}
\medskip
\centerline{and Rolling Closed String Tachyon}}}
\centerline{Yasuaki Hikida\foot{E-mail:
hikida@phya.snu.ac.kr} and Tadashi Takayanagi\foot{E-mail:
takayana@bose.harvard.edu}}

\bigskip\centerline{\it ${}^1$School of Physics \& BK-21 Physics Division,
Seoul National University}
\centerline{\it Seoul 151-747, Korea}
\medskip\centerline{\it ${}^2$Jefferson Physical Laboratory,
Harvard University}
\centerline{\it Cambridge, MA 02138, USA}

\vskip .3in
\centerline{\bf Abstract}

\vskip .1in We investigate the $SL(2,\blackb{R})/U(1)$ WZW model
with level $0<k<2$ as a solvable time-dependent background
in string theory. This model is expected to be dual to the one
describing a rolling closed string tachyon with a
time-like linear dilaton. We examine its exact metric and
minisuperspace wave functions. Two point functions and the
one-loop vacuum amplitude are computed and their relation to the
closed string emission is discussed.
Comparing with the results from the minisuperspace approximation,
we find a physical interpretation of our choice to continue the
Euclidean model into the Lorentzian one.
Three point functions are also examined.

\noblackbox

\Date{}

\writetoc

\ftno = 0

\newsec{Introduction}

The open string tachyon condensation has been widely investigated
up to now, and we obtain much insights on the decay process of
unstable D-branes. However, it is very difficult to analyze closed
string tachyon condensation except for the successful examples of
the localized closed string tachyons initiated in \APS\ on
non-supersymmetric orbifolds\foot{For recent developments on this
subject see reviews \HMT\ and references therein.}.
This is because of the lack of tractable conformal field theories
for closed string tachyon condensation, which would change the
background itself. Nevertheless, there have been some attempts to
describe the homogeneous rolling tachyon condensation of closed
strings by using a Lorentzian continuation of the bulk Liouville
theory \Ma \ST \SC\ (we call this the time-like Liouville theory).
This idea is inspired by a similar process of open string tachyon
condensation \Sen\ and the related time-like boundary Liouville theory \SO
\GS \LNT. The closed string tachyon condensation in two
dimensional string theory was recently discussed in \KSMC
\KSMCC\ by applying the matrix model description.

It is also very interesting to study inhomogeneous closed string
tachyon condensations such as a tachyon kink, just like the open
string case. The simplest example may be obtained from a
Lorentzian continuation of sine-Liouville theory\foot{One can find
in \KuHu\ earlier discussions on the condensation of a tachyon kink
using the Euclidean sine-Liouville theory or equally a non-critical
string with a tachyon kink.}. The usual sine-Liouville
theory is defined by the action (we always assume $\al=1$ below)
 \eqn\sinl{S=\int
dz^2\left[ \de X\bar{\de} X+\de \phi\bar{\de}\phi+\lambda
e^{b\phi}\cos\left(\f{X}{R}\right)\right],} where $\phi$ is a
Liouville field with background charge $Q$ (such that the string
coupling $g_s=e^{-Q\phi}$), and $X$ is a free boson compactified
on a circle with radius $R$. The total central charge is
$c=2+6Q^2$ and the conformal dimension of the primary field
$e^{b\phi}$ is $\Delta_b = -\f{1}{4}b(b+2Q)$.

Performing the Lorentzian continuations $\phi=-iX^0$ and
$b=i\beta$ ($X^0$ and $\beta$ are real), we obtain
\eqn\timel{S=\int dz^2\left[ \de X\bar{\de} X-\de
X^0\bar{\de}X^0+\lambda e^{\beta
X^0}\cos\left(\f{X}{R}\right)\right].} Obviously, this action
describes the condensation of a tachyon kink on the circle. Even
before the Lorentzian continuation, sine-Liouville theory may not be
easily solved exactly for generic parameters. However, at the
specific value of parameters \eqn\paramm{Q=\f{1}{\s{k-2}},\qquad
R=\f{1}{\s{k}},\qquad b=-\f{1}{Q}=-\s{k-2},} we can solve it by
making use of the Fateev-Zamolodchikov-Zamolodchikov (FZZ) duality
(see e.g. \GK \KKK \FH \GKK), which relates the theory to the
$SL(2,\blackb{R})/U(1)$ WZW model with level $k$ (and radius
$\ti{R}=\f{1}{R}=\s{k}$). This is a sort of T-duality and can be
proved explicitly in the case of ${\cal N}=2$ Liouville theory
\HK. If we extend this duality relation to the region $0<k<2$
analytically, then we can analyze the previous
inhomogeneous closed string tachyon condensation \timel\ using the
relation to the $SL(2,\blackb{R})/U(1)$ WZW model. The latter model is
much more useful because of not only technical reasons (e.g., of
computing correlation functions) but also intuitive
understandings. Indeed we can find an exact metric in the dual
picture as we will see later.

Motivated by these facts, in this paper we mainly study the
$SL(2,\blackb{R})/U(1)$ theory extended to the region $0<k<2$,
where $R$ is real and $Q$ and $b$ are pure imaginary. As in the
sine-Liouville theory, all of the physical parameters become real
by regarding the linear dilaton field (in a free field
realization) as a time coordinate. This coset model itself, of
course, gives also an interesting solvable time-dependent
background\foot{See \TV \KL \LU \BLW \CKR\ for earlier
discussions on cosmological interpretations of
$SL(2,\blackb{R})/U(1)$ model. Also refer to e.g. a review \CO\
for recent developments of time-dependent backgrounds in string
theory.} in string theory.

The rest of the paper is organized as follows.
In section 2 we examine the geometry of the time-dependent background
using the exact metric with stringy corrections. In section 3 we apply the
minisuperspace approximation to our model and
examine the wave functions. In section 4 we compute the exact
two point function and discuss its relation to the closed string pair
creation. We also find the exact three point function.
In section 5 we analyze the decay of the vacuum by using
the one-loop torus partition function.
In section 6 we comments on the relation to the rolling closed
string tachyon, and in section 7
we summarize the conclusions and discuss future problems.

\newsec{Exact Time-Dependent Geometry}
Originally, the $SL(2,\blackb{R})/U(1)$ WZW model\foot{There are two
types of $U(1)$ symmetry in the $SL(2,\blackb{R})$ WZW model; one
is time-like and the other is space-like. We obtain the Euclidean
(Lorentzian) geometry by gauging time-like (space-like) $U(1)$,
which corresponds to the Euclidean (Lorentzian) 2d black hole.
In this paper we mean the $SL(2,\blackb{R})/U(1)$ model as the one
gauged by time-like $U(1)$.} was investigated
to describe the Euclidean two dimensional black hole in the
context of string theory \Wi \MSW . The level of the WZW model is
labeled by $k$, and the exact geometry of the model is known
including $\alpha '$ correction \DVV \TSQ. Roughly speaking, the
quantum correction shifts the level $k \to k-2$, and the Euclidean
signature of the metric requires $k>2$. If we go beyond this bound
into $0<k<2$, then we obtain an interesting Lorentzian model\foot{The
Lorentzian $SL(2,\blackb{R})/U(1)$ model with negative level $k<0$ also
leads to a different Lorentzian geometry, which was
investigated in \CKR. We will not mainly discuss the case
since we are motivated by the dual
background with the inhomogeneous closed string tachyon
condensation as discussed in section 1. See also section 7 on
this point.} as we will see below.

\subsec{Exact Geometry of $SL(2,\blackb{R})/U(1)$ Model with $0<k<2$}

In order to see the geometry given by the $SL(2,\blackb{R})/U(1)$
model, we should use the metric with $\alpha'$ corrections since
the shift $-2$ of the level $k$ is relevant in the action \timel\
and the parameters \paramm .
Fortunately, the $\alpha '$ exact metric of the model can be found in
\DVV \foot{For a general $\al$ the metric $ds^2$ is modified by
overall factor $\al$. If we take the large $k$ limit, then the metric
reduces to the familiar two dimensional black hole metric
$ds^2=k\al (dr^2+\tanh^2 r d\theta^2)$ with $g_s^2= 1/\cosh^2 r$.}
as (we always assume $\al=1$ below)
\eqn\met{ds^2=(k-2)\left(dr^2+\f{\tanh^2 r}{1-\f{2}{k}\tanh^2
r}d\theta^2\right),}
where $\theta$ is compactified as
$\theta\sim \theta+2\pi$ and $r$ is restricted to $r\geq 0$.
The string coupling is given by
\eqn\gs{
g_s^2 = e^{2\Phi}=\f{1}{\cosh r\s{\cosh^2 r-\f{2}{k}\sinh^2 r}}.}
This metric represents a cigar-like geometry
with the asymptotic radius $\ti{R}=\s{k}$
, and, in particular for
$k=9/4$, the critical string model describes the two dimensional
(Euclidean) black hole \Wi \MSW.
The exactness of the metric was proven in \TSQ\ by
noticing the quantum corrections of the levels of the coset WZW model
(see also \Pi \PS\ for further discussions).

In order to define a consistent critical string theory, we have to
add other conformal field theories with total central charge
$c=24-6/(k-2)$. We can assume, for example, that the additional
sector consists of 24 scalar bosons $X_2,X_3,\ddd,X_{25}$ with a
non-trivial linear dilaton. There are many other candidates for
such an additional sector, but the sector does not play an
important role in our discussions. Because the total system always
preserves conformal symmetry, the backgrounds with any value of $k$
correspond to a series of exact solutions to the $\alpha'$
corrected Einstein-dilaton gravity for bosonic string theory.
Therefore, it must be possible to obtain an exact classical
solution continuing into the region $0<k<2$.

Using the Lorentzian continuation of the level $k$, we obtain the
following metric and dilaton\foot{We use the fact that the dilaton
value is determined only up to a constant shift.} (we also replace
$r$ with $t$ to make its signature clear)
 \eqn\metl{ \eqalign{ ds^2&=-\ap dt^2+\ap \f{\tanh^2
t}{\f{2}{2-\ap}\tanh^2 t-1}d\theta^2,\cr
 g_s^2 &= \f{1}{\cosh t \s{ \f{2}{2-\ap}\sinh ^2 t
- \cosh ^2 t}},}}
where we defined
\eqn\alp{\alpha\equiv 2-k>0.}
This metric and the dilaton have a space-like singularity at
$t=\pm t^*$ (curvature is divergent), where $t^*>0$
is a solution to $\f{2}{2-\ap}\tanh^2 t-1=0$.
Therefore, we use a region $t^*<t<\infty$ to define the spacetime,
though we could use the other region $-\infty<t<-t^*$,
which can be obtained by replacing $t\to -t$.
The restriction of the coordinate is natural
since the original metric is defined only for $r<0$ or $r>0$.

Let us see the asymptotic behaviors of the metric.
At the late time $t \gg 1$ the background approaches to a
flat spacetime (with the asymptotic radius $\ti{R}=\s{2-\ap}$)
with a time-like linear dilaton $g_s\sim e^{-t}$.
Near the starting point $t^*$, where the metric and the dilaton diverge,
we find the approximation of the metric and the dilaton
\eqn\singu{ds^2=-\ap dt^2+\left(\f{2-\ap}{2}\right)^{\f{3}{2}}
\f{d\theta^2}{t-t^*}, \qquad g_s^2=\s{\f{\ap}{(4-2\ap)(t-t^*)}}.}
In conclusion, this time-dependent background describes a spacetime
which starts with a singularity and contracts into a weakly coupled
flat spacetime with a time-like linear dilaton.

So far we have discussed the geometry described by the string
metric \metl . It is also useful to examine other metrics such as the
Einstein metric or the one observed by a D0-brane.
Each of them is defined by $(ds')^2=(g_s)^{a}ds^2$, where
the constant $a$ is $-1/6$ for the Einstein metric and
$-2$ for the D0-brane metric.
It is easy to see that the both metrics again
have the similar singularity at $t=\pm t^*$.

\subsec{Exact Geometry of T-dual Model}

In the $SL(2,\blackb{R})/U(1)$ model, we can find the metric after
the T-duality along the compactified circle including $\alpha '$
correction. For the Euclidean case $k>2$, the exact metric for the
T-dual model is given in \DVV \eqn\exmtt{\eqalign{ ds^2&=
(k-2)\left(dr^2+\f{d\ti{\theta}^2}{\tanh^2r-\f{2}{k}} \right), \cr
g_s^2&=\f{1}{\sinh r\s{\sinh^2 r-\f{2}{k}\cosh^2 r}},}} where
$\ti{\theta}$ is the dual coordinate of $\theta$ and is
compactified as $\ti{\theta}\sim \ti{\theta}+2\pi/k$. As before we
obtain the exact time-dependent background with $0<k<2$ using the
Lorentzian continuation as \eqn\timedmd{\eqalign{ ds^2&= -\ap\
dt^2+\f{\ap\ d\ti{\theta}^2}{\f{2}{2-\ap}-\tanh^2 t}, \cr
g_s^2&=\f{1}{\sinh t\s{\f{2}{2-\ap}\cosh^2 t- \sinh^2 t}}.}}
Interestingly, the metric has no singularity in contrast to the
previous metric. The radius of the compactified direction starts
from $R_{0}=\s{\f{\ap}{2(2-\ap)}}$ at the beginning $t=0$,
and it approaches to $R_{\infty}=\f{1}{\s{2-\ap}}$ at the late
time $t=\infty$. Since the string coupling diverges at $t=0$, we
can restrict\foot{We can use $t<0$ by
replacing $t\to -t$.} the time coordinate to $t>0$.
This is roughly consistent with the
structure of the coordinate system \metl\ before the T-duality.

At the late time $t \gg 1$, the background approaches to a flat
spacetime with a time-like linear dilaton $g_s\sim e^{-t}$.
Naively, the absence of singularity at $t=\pm t^*$ does not seem
consistent with the geometry before the T-dual transformation.
Even though we do not have a complete answer to this question, it
is probable that the singularity at $t=\pm t^*$ in the original
metric is actually not true one in string theory. This observation
may be deduced from the analysis of the minisuperspace
approximation in the next section. The wave function does not have
a singular behavior at $t=\pm t^*$ and we can indeed reproduce
the exact two point function computed later by just assuming the
boundary condition at $t=0$ neglecting\foot{Note also that the
singularity of the dilaton coincides with that of the metric. It is
possible that the both are canceled with each other.} the
`singularity' at $t=\pm t^*$. Notice also that such a `discrepancy'
already exists in the Euclidean theory.

Although the string metric of this background is smooth
everywhere, the singularity $t=0$ appears in the dilaton field.
Therefore, the other kinds of metrics (Einstein
metric and D0-brane metric), discussed in the end of the
previous subsection, become singular at $t=0$. Rewriting such
metrics in a form $ds'^2=-dt'^2+a(t)^2 d\theta^2$, we obtain the
behavior $a(t)\sim t^b$ near $t=0$, where the constant $b$ is
given by $b=\f{1}{25}$ for the Einstein metric and $b=\f{1}{3}$
for the D0-brane metric. This implies that the two dimensional
spacetime gives an expanding behavior near the singularity $t=0$
for any kinds of metric considered so far. This is T-dual
to the previous contracting spacetime.

\newsec{Minisuperspace Limit of the Time-Dependent Model}

Before moving to the exact conformal field theory analysis, we
deal with the minisuperspace limit of string theory on \metl \ or
\timedmd . Finding the eigenfunction of the minisuperspace
Hamiltonian means solving a Klein-Gordon equation for a scalar
field\foot{For the minisuperspace analysis in Euclidean
$SL(2,\blackb{R})$ or $SL(2,\blackb{R})/U(1)$ model see e.g. \DVV
\TSM \SHM.}. We first concentrate on the case with \timedmd \
because the rolling tachyon model \timel\ is FZZ dual (T-dual)
to the $SL(2,\blackb{R})/U(1)$ model and hence directly equivalent to the
T-dual model \timedmd . {}From the exact metric \timedmd ,
we find that the Laplacian $\Delta=-\f{e^{2\Phi}}{\s{-g}}
\de_{\mu}(e^{-2\Phi}\s{-g}g^{\mu\nu}\de_{\nu})$
is
\eqn\laplacian{\eqalign{
 \Delta &= \frac{1}{\alpha}
 \left( \partial_t^2 + 2 \coth 2t \partial_t
 +
 \left( \tanh^2 t -\frac{2}{2-\alpha} \right)
         \partial_{\ti{\theta}}^2 \right)\cr
 &= \frac{1}{\alpha} \left(4 \partial_y\left( y(1+y) \partial_y\right)
    + \left( - \frac{1}{1+y} + 1
    - \frac{2}{2-\alpha} \right) \partial_{\ti{\theta}}^2\right)}
} with $y=\sinh^2 t$. The eigenfunctions of the Laplacian are
given by linear combinations of two independent solutions, the
regular one (note that $k n'\in k \blackb{Z}$ due to the
periodicity $\ti{\theta}\sim \ti{\theta}+2\pi /k$) \eqn\sol{
 \Phi^{(1)} = e^{i k n' \ti{\theta}} (1+y)^{\nhk}
 F\left(-j+\nhk, 1+j+\nhk,1;-y\right),
} and the logarithmic one \eqn\logsol{\eqalign{
 \Phi^{(2)} &= e^{i k n' \ti{\theta}} (1+y)^{\nhk} \Bigl(
 F\left(-j+\nhk,1+j+\nhk,1;-y \right) \ln (-y) \cr
 & +\sum_{l=1}^{\infty} (-y)^l
 \frac{(-j+\nhk)_l (1+j+\nhk)_l}{(l!)^2}
  \{ \psi (-j  + \nhk + l) - \psi (-j  + \nhk) \cr
 & \qquad \qquad
 + \psi (1 +j + \nhk + l) - \psi (1 +j + \nhk) - 2 \psi (1 + l)
 + 2 \psi (1) \} \Bigr),
 } }
where we have defined
\eqn\al{ (a)_l = \frac{ \Gamma (a + l)}{\Gamma (a)} ,
 \qquad \psi(x) = \frac{d}{dx} \ln \Gamma (x) .}
The eigenvalues are\foot{The label $j$ should be written in terms of
closed string oscillators and momenta of other directions if a critical
string theory is considered.}
\eqn\eigenvalue{\Delta = \frac{4j(j+1)}{\alpha} + \frac{{n'}^2}{(2-\alpha)}.}
Here we assume the boundary condition that the wave function should
be smooth at $t=0$. Then the first solution $\Phi^{(1)}$ is selected
since the second solution $\Phi^{(2)}$ diverges at $t = 0$.
After normalizing as\foot{We can show the self-adjointness of the
minisuperspace Hamiltonian with the eigenfunction by following \TSM .
See \mini \minitwo\ for the analogue of the time-like Liouville theory.
Our results, which are also consistent with the exact CFT result, almost
corresponds to the specific case $\nu_0=0$ in \mini.}
\eqn\invac{ \Psi  = \frac{\Gamma (1+j +\nhk)\Gamma (1+j -\nhk)}
  {\Gamma (2j+1)} \Phi^{(1)} ,}
we find its behavior for large $y$ ($\sim (e^t/2)^2 $)
\eqn\asympt{ \Psi
 \sim  e^{i k n' \tilde{\theta}} y^{j} +
    c^j_{n'} e^{i k n' \tilde{\theta}} y^{-1-j}
  ,}
with
\eqn\reflectiond{c_{n'}^{j}
 = \frac{\Gamma(1+j+\nhk) \Gamma(1+j-\nhk)\Gamma (-2j-1)}
        {\Gamma(-j+\nhk) \Gamma(-j-\nhk) \Gamma (2j+1)}.}
Namely, for $j=-\f{1}{2} +i \f{\sqrt\alpha}{2}\omega\
 (\omega \in \blackb{R})$,
we can interpret the eigenfunction as a linear combination of
positive and negative frequency modes. The ``reflection
coefficient'' \reflectiond\ can be regarded as the two point
function $\la e^{i\omega X_0} e^{i\omega X_0} \lb$ in our
Lorentzian theory. This fact is well-known in the usual Liouville
theory or the Euclidean $SL(2,\blackb{R})/U(1)$ model (see e.g. \ZZL \AA
\AB). In the presence of the wall due to the screening operator,
waves are reflected back into the weakly coupled region as
$\psi(\phi)\sim e^{ip\phi}+S(p)e^{-ip\phi}$. Then it is obvious
that the (non-trivial) two point function is given by $\la
e^{ip\phi} e^{ip\phi} \lb=S(p)$. For our time-like case we
extend this result by the Lorentzian continuation. As we will see
in the next section, our result in the minisuperspace model
is consistent with the one of the two point functions from the exact
conformal field theory computations. Furthermore we later discuss
that the reflection
coefficient $c^j_{n'}$ represents the closed string pair production
as in the time-like Liouville theory \SO \GS \Ma \ST .

We can also apply the minisuperspace analysis to the original exact
classical background \metl. The Laplacian takes the similar form
\eqn\laplaciano{\eqalign{
 \Delta &= \frac{1}{\alpha}
 \left( \partial_t^2 + 2 \coth 2t \partial_t
 + \left( \coth^2 t -\frac{2}{2-\alpha} \right) \partial_\theta^2 \right)\cr
 &= \frac{1}{\alpha} \left(4 \partial_y y(1+y) \partial_y
    + \left(\frac{1}{y} + 1
    - \frac{2}{2-\alpha} \right) \partial_{\theta}^2\right)}
}
with $y=\sinh^2 t$. There are two independent solutions which has the
eigenvalue \eigenvalue\ (with $n' \leftrightarrow n$)
of the Laplacian \laplaciano\ as in the previous case.
Assuming the non-singular boundary condition at $t=0$ again,
we get the solution ($n\in \blackb{Z}$)
\eqn\sol{
 \Phi^{(1)} = e^{i n \theta} y^{\nh}
 F\left(-j+\nh, 1+j+\nh,1+|n|,-y\right).
}
{}From this we can find the ``reflection coefficient'' (two point function)
\eqn\reflection{d_{n}^{j}
 = \frac{(\Gamma(1+j+\nh))^2 \Gamma (-2j-1)}
        {(\Gamma(-j+\nh))^2 \Gamma (2j+1)},}
and this is consistent with the exact results discussed later. Notice
that in this argument the singularity of the
metric and dilaton at $t=\pm t^*$
seems not important as mentioned in the previous section.

\newsec{Exact Analysis of the Time-Dependent String Theory}

\subsec{Two Point Function} We would like to turn to the exact
conformal field theory analysis. For the Euclidean model of
$SL(2,\blackb{R})/U(1)$ with $k>2$, the two point function was
computed for the primaries $V_{j,m,\bar{m}}$, in \GK\foot{It
might be useful to observe that the two point function
is invariant under $(-m,\bar m) \leftrightarrow (m,- \bar m)$
when comparing to the minisuperspace results.}
\eqn\twope{\la V_{j,m,\bar{m}}V_{j,-m,-\bar{m}} \lb
=(k-2)(\nu(k))^{2j+1}
\f{\Gamma\left(1-\f{2j+1}{k-2}\right)\Gamma(-2j-1)\Gamma(1+j-m)\Gamma
(1+j+\bar{m})}{\Gamma\left(\f{2j+1}{k-2}\right)
\Gamma(2j+2)\Gamma(-j-m)\Gamma (-j+\bar{m})},} where
\eqn\nuk{\nu(k)=\f{\Gamma\left(1+\f{1}{k-2}\right)}
{\pi\Gamma\left(1-\f{1}{k-2}\right)}.} When we shift the coupling
constant, the function $\nu(k)$ scales as $\nu\sim (g_s)^{-2}$.
The labels $(m,\bar m)$ take the value
\eqn\mbarm{m=\frac{n+kn'}{2},\qquad \bar m = -\frac{n-kn'}{2},\qquad
 n,n' \in \blackb{Z}.}
In the dual sine-Liouville model, the tachyon vertex operators
$V_{j,m,\bar{m}}$ correspond to the operators
\eqn\vert{V(p_L,p_R,\omega)=e^{-\f{X^0}{\s{\ap}}}e^{i\omega
X^0+ip_LX_L+ip_R X_R},} where the relation between $(j,m,\bar{m})$
and $(\omega,p_L,p_R)$ is given by
\eqn\rel{j=-\f{1}{2}+i\f{\s{\ap}}{2}\omega,\qquad m=\f{\s{k}}{2}p_L,\qquad
\bar{m}=-\f{\s{k}}{2}p_R.} Then the two point function for
$0<k<2$ and $Q=\f{i}{\s{\ap}}$ is given by \eqn\twol{\eqalign{ & \la
V(p_L,p_R,\omega) V(-p_L,-p_R,\omega)\lb \cr
&=-(\nu(k))^{i\s{\ap}\omega}
\f{\Gamma\left(i\f{\omega}{\s{\ap}}\right)
\Gamma\left(-i\s{\ap}\omega\right)
\Gamma\left(\f{1}{2}+i\f{\s{\ap}}{2}\omega-\f{\s{k}}{2}p_L\right)
\Gamma\left(\f{1}{2}+i\f{\s{\ap}}{2}\omega-\f{\s{k}}{2}p_R\right)}
{\Gamma\left(-i\f{\omega}{\s{\ap}}\right)
\Gamma\left(i\s{\ap}\omega\right)
\Gamma\left(\f{1}{2}-i\f{\s{\ap}}{2}\omega-\f{\s{k}}{2}p_L\right)
\Gamma\left(\f{1}{2}-i\f{\s{\ap}}{2}\omega-\f{\s{k}}{2}p_R\right)},}}
where \eqn\nukk{\nu(k)=\f{\Gamma(1-1/\ap)}{\pi\Gamma(1+1/\ap)}.}
Applying the FZZ duality to the time-dependent case,
this result should also give the
two point function of the time-dependent sine-Liouville theory
defined by \timel. We just suppressed the dependence on the parameter
$\lambda$, which scales as $(g_s)^{\ap}$.

The poles in the two point function \twol\ can be understood as
follows. The poles  $\omega= - i n /\s{\ap}$ ($n \in \blackb{Z}$)
of $\Gamma (-i\s{\ap}\omega)$ correspond to the
insertion of the screening operator $\sim
e^{-2Q\phi}=e^{-2X^0/\s{\ap}}$ to the correlators in the free
field representation of $SL(2,\blackb{R})/U(1)$ model. On the
other hand, the poles $\omega=i n \s{\ap}$
coming from $\Gamma (i\omega/\s{\ap})$ can be explained as the
contributions from the screening operator\foot{Notice that only
even power of the screening operator contributes due to the
cosine combination.} or equally the tachyon perturbation
$T(x_0,x)=\lambda e^{\s{\ap}X^0} \cos(\s{2-\ap}\ X)$ in the action
\timel. All other poles are reproduced\foot{The part obtained from
the minisuperspace analysis can be also reproduced from a group
theoretical consideration as $\int d^2 x x^{j+m} \bar x^{j + \bar
m} |x-1|^{-4(1+j)}$.} from the minisuperspace approximation done
in section 3. Indeed we find that the minisuperspace results
\reflectiond\ and \reflection\ are exactly the same as \twol\ for
non-zero winding and momenta except the first factor $\Gamma
(i\omega/\s{\ap})$, respectively. Missing this factor is very
natural because it is related to the dual sine-Liouville theory
and this duality occurs only in string theory and not in a field
theory limit.

An interesting limit would be $\ap=2$ ($k=0$). This should lead to
a purely time-like Liouville theory $T=\lambda e^{\s{2}X^0}$ as
the cosine part becomes trivial. If we compute the two point
function in this limit (we can set $p_L=p_R=0$), we get
\eqn\twops{ \la V(\omega) V(\omega)\lb
=-(2\pi)^{-i\s{2}\omega}\f{\Gamma\left(\f{1}{2}+\f{i\omega}{\s{2}}\right)}
{\Gamma\left(\f{1}{2}-\f{i\omega}{\s{2}}\right)}.} Indeed this is
the same as that computed in the time-like Liouville theory \ST
\eqn\twotl{\la V(\omega)
V(\omega)\lb=-\left(\f{\pi\lambda\gamma\left(-\f{1}{2}\right)}
{4}\right)^{-\s{2}i\omega}
\f{\Gamma\left(\f{1}{2}+\f{i\omega}{\s{2}}\right)}
{\Gamma\left(\f{1}{2}-\f{i\omega}{\s{2}}\right)},} after choosing
a specific $\lambda$. It might be also useful to note that
in this limit the metric \timedmd\ becomes flat and the
string coupling $g_s^2$ is proportional to $\f{1}{\sinh(2t)}$.
This represents the backreaction to the dilaton field (and
no backreaction to the metric) in the time-like
Liouville theory or equally the homogeneous closed string
tachyon condensation.

\subsec{Closed String Pair Production}

In general a non-zero two point function $\la e^{-i\omega
X_0}e^{-i\omega X_0}\lb$ implies that a pair of closed string with
energy $\omega$ can be produced in that background as discussed in
detail in the context of the time-like (boundary) Liouville theory
\SO \GS \Ma \ST. Thus the non-zero two point function \twol\ shows
that this interesting phenomenon occurs also in our case. Since
the production rate for each string mode is expected to be
proportional\foot{Here we have first taken the absolute valued
square of the two point function and extends it to the Lorentzian
region. Instead, if we first take the Lorentzian continuation,
then we have either $|\la e^{-i\omega X_0}e^{-i\omega
X_0}\lb|^2=1$ (e.g. for all cases in the region $0<k<1$) or $|\la
e^{-i\omega X_0}e^{-i\omega X_0}\lb|^2=e^{-2\pi\s{\ap}\omega}$
depending the value of the level $k$.
Even in the latter case, it is overwhelmed by the exponentially large
density of state and thus the closed string production is always
exponentially divergent.} to $|\la e^{-i\omega X_0}e^{-i\omega
X_0}\lb|^2=1$, the total closed string production can be estimated
by summing over the Hagedorn tower of massive states with the
density of state $\rho(\omega)$ \eqn\hagec{\int d\omega
\rho(\omega)|\la e^{-i\omega X_0}e^{-i\omega X_0}\lb|^2 \sim \int
d\omega\ e^{4\pi \s{1+\f{1}{4\ap}}\omega}.} This is exponentially
divergent and shows that there will be a huge backreaction due to
the pair creation. The similar divergence was also observed in the
time-like bulk Liouville theory \Ma \ST.

\subsec{Three Point Function}

We can also compute the three point function of the
$SL(2,\blackb{R})/U(1)$
time-dependent model by extending the exact result \GK\ for
$k>2$. It is given by
\eqn\threep{\la
V_{j_1,m_1,\bar{m_1}}V_{j_2,m_2,\bar{m_2}} V_{j_3,m_3,\bar{m_3}}
\lb =D(j_1,j_2,j_3)\cdot F(j_i,m_i,\bar{m_i})\cdot
\delta^2(\sum_{i=1}^3 m_i),}
where $F(j_i,m_i,\bar{m_i})$ is a group theoretical coefficient
\eqn\cg{\eqalign{ &
F(j_i,m_i,\bar{m_i}) =\left(\prod_{i=1}^{3}\int dx_i^2
 x_i^{j_i+m_i}\bar{x_i}^{j_i+\bar{m_i}}\right) \cr & \qquad \cdot
|x_1-x_2|^{2(j_3-j_1-j_2-1)}
|x_2-x_3|^{2(j_1-j_2-j_3-1)}|x_3-x_1|^{2(j_2-j_3-j_1-1)},}} which
is not possible to express in a simple way generically. For our
time-dependent background $0<k<2$, we will substitute the values
$j_i=-1/2+i\s{\ap}\omega_i/2$ as before for the energy $\omega_i$
state. The non-trivial physics is included in the factor
$D(j_1,j_2,j_3)$, which was computed in \AA \AB \Is\ as\foot{Notice
that we are using a different convention of $j$ and it is related
to that of \AA \AB\ via $j\to -j-1$.}
\eqn\three{\eqalign
{&D(j_1,j_2,j_3) = \f{k-2}{2\pi^3}\nu(k)^{j_1+j_2+j_3+1} \cr
&\cdot \f{G(-j_1-j_2-j_3-2)
G(j_3-j_1-j_2-1)G(j_2-j_1-j_3-1)G(j_1-j_2-j_3-1)}
{G(-1)G(-2j_1-1)G(-2j_2-1)G(-2j_3-1)}.}} The function $G(j)$ is
defined by
\eqn\gj{G(j)=\ti{b}^{-\ti{b}^2j(j+1+\ti{b}^{-2})}
\Upsilon^{-1}(-\ti{b}j),\qquad \ti{b}=\f{1}{\s{k-2}}\,(=Q)}
in terms of the function
$\Upsilon_{\ti{b}} (x)$ introduced in \ZZL.
It might be convenient to express \three\ as
\eqn\threeu{\eqalign {&D(j_1,j_2,j_3) =
\f{1}{2\pi^3}\nu(k)^{j_1+j_2+j_3+1}
\ti{b}^{-2\ti{b}^2(j_1+j_2+j_3+1)-1} \cr & \!\! \cdot
\f{\Upsilon(\ti{b})\Upsilon((2j_1+1)\ti{b})
\Upsilon((2j_2+1)\ti{b})\Upsilon((2j_3+1)\ti{b})}
{\Upsilon((j_1+j_2+j_3+2)\ti{b}) \Upsilon((j_1+j_2-j_3+1)\ti{b})
\Upsilon((j_3+j_1-j_2+1)\ti{b})\Upsilon((j_2+j_3-j_1+1)\ti{b})}.}}
In order to obtain the three point function in our case,
we perform the Lorentzian continuation to the region
$0<k<2$.

Contrary to the two point function, it is not straightforward to
continue to the region $0<k<2$ since the function
$\Upsilon_{\ti{b}}(x)$ is not originally defined for imaginary
values of the parameter $\ti{b}$ (see \paramm).
Fortunately we can accomplish the continuation procedure
in a similar way as in \SC\ done for the time-like Liouville theory.
The result is
\eqn\threept{\eqalign {&
D(\omega_1,\omega_2,\omega_3) = -\f{1}{2\pi^3}
(\nu(k))^{-\f{1}{2}+\f{i\s{\ap}(\omega_1+\omega_2+\omega_3)}{2}}\cdot
e^{-i\f{\pi}{\ap}}\cdot
(\ap)^{-\f{i}{2\s{\ap}}(\omega_1+\omega_2+\omega_3)+\f{1}{2\ap}+\f{3}{2}}
\cr & \qquad \cdot
\f{H(\f{i}{\s{\ap}})H\left(-2\omega_1-\f{i}{\s{\ap}}\right)
H\left(-2\omega_2-\f{i}{\s{\ap}}\right)
H\left(-2\omega_3-\f{i}{\s{\ap}}\right)}
{H\left(-\omega_1-\omega_2-\omega_3\right)
H\left(\omega_3-\omega_1-\omega_2\right)
H\left(\omega_2-\omega_1-\omega_3\right)
H\left(\omega_1-\omega_2-\omega_3\right) },}}
where we define
\eqn\funh{
H(\omega)=\theta_1\left(-\f{1}{2}+\f{i\s{\ap}\omega}{2},
\ap\right)\cdot Y_{\f{1}{\s{\ap}}}(\omega).} The function
$Y_{\beta}(x)$, originally introduced\foot{In the
notation of \SC\ this is written as $Y_{\f{1}{\s{\ap}}}
\left(-\f{i}{\s{\ap}}j\right)$.} in \SC,
is written in the form
\eqn\funcy{ \log
Y_{\f{1}{\s{\ap}}}(\omega)=const.+\int_0^{\infty}\f{d\tau}{\tau}
\left[e^{-\tau}\f{(\omega+i\s{\ap})^2}{4} -\f{\sin^2
\left(\f{(\omega+i\s{\ap}) \tau}{4}  \right)}
{\sinh\left(\f{\tau}{2\s{\ap}}\right)\sinh\left(\f{\s{\ap}\tau}{2}\right)}
\right].}
Notice that the above formula becomes singular at
$\ap=1$ (or $k=1$) due to the theta function in \funh.
The singularity can be analyzed in the same way as \SC .

\newsec{Torus Amplitude and Decay Rate of the Vacuum}

Another quantity related to the closed string production may be
(the imaginary part of) the one-loop vacuum amplitude\foot{In the
case of open string rolling tachyon, there is a definite relation
that the imaginary part of the cylinder amplitude and the
square root of the amplitude of closed string emission are the same
\LLM \KLMS . We expect a similar relation between the torus amplitude
and the rate of the closed string production (or two point function),
although the explicit relation is unclear.
Indeed our result in section 5.3
implies there is such a relation between them.}.
In the next
subsections we compute the torus amplitude in the model with
$SL(2,\blackb{R})/U(1) \times (D-2)$ free bosons. We find that the
torus amplitude diverges, which may be rewritten as the imaginary
part by an analytical continuation. We interpret the imaginary part as
the decay rate of the vacuum (see e.g. \Marcus \WW). In the other
subsections we see that there are two types of origin of the
divergence. One is the emission of the tachyonic modes just like
the ordinary bosonic string theory on flat spacetime. The other is
due to the poles of the density of states, and we propose that
the divergence is
interpreted as the emission of the closed strings with high
energy as discussed before.

\subsec{One-loop Amplitude in the Time-Dependent String Theory}

First let us assume that the time-dependent string theory consists
of $SL(2,\blackb{R})/U(1)$ with level $k > 2$ and $D-2 \,(\,= 24 -
6/(k-2))$ free bosons\foot{We choose a different additional sector
from the one in section 2 just for convenience.}.
Then, it is straightforward to write down
the modular invariant vacuum amplitude computed from the
path-integral formulation \HPT\ (see also \Gaw \MOS)
\eqn\partt{\eqalign{Z &=2\s{k(k-2)}\int_{\cal F} \f{d\tau
d\bar{\tau}}{\tau_2} \int^1_{0} ds_1 ds_2 \cr
&\sum_{w,m=-\infty}^{\infty} \sum_i
q^{h_i}\bar{q}^{\bar{h}_i}e^{4\pi\tau_2(1-\f{1}{4(k-2)})
-\f{k\pi}{\tau_2}|(s_1+w)\tau-(s_2+m)|^2+2\pi\tau_2 s_1^2} \cr
&\f{1}{|\sin (\pi(s_1\tau-s_2))|^2}
\prod_{r=1}^{\infty}\f{|1-e^{2\pi ir\tau}|^4}{|1-e^{2\pi
ir\tau-2\pi i(s_1\tau-s_2)}|^2 |1-e^{2\pi ir\tau+2\pi
i(s_1\tau-s_2)}| ^2}.}} We denote $h_i,\bar{h}_i$ as the conformal
dimensions of $D-2$ free bosons and $q=\exp (2 \pi i \tau)$ $(\tau
= \tau_1 + i \tau_2)$ as the moduli of the torus. We perform the
integral of $\tau$ in the fundamental region ${\cal F}$.

For the value $0<k<2$ considered now, we should perform the
Lorentzian continuation of the above amplitude\foot{Notice
that the coefficient includes the imaginary factor
$\s{k-2}=-i\s{\ap}$ after the continuation.
This implies that the spacetime is truly Lorentzian just as the volume
factor in the case of the flat spacetime becomes $iV_{26}$
after the Wick rotation.}.
As was done in the Euclidean case \HPT, we can perform the
Poisson resummation with respect to $m$ and introduce an extra
integration of the continuous valuable $s$ (`energy') in order to get
the partition function in terms of the operator formalism.
One important point in our Lorentzian case is that
we should deform the contour of the integral of $s$ into
the imaginary axis ($\pi/2$ rotation) as follows
\eqn\intt{\f{1}{2i}\s{\f{\ap}{\tau_2}}
=\int^{i\infty}_{-i\infty}\ ds\ e^{4\pi\tau_2s^2/\ap}.}
To make this explicit we define $s=i\ti{s}$.
The partition function in the end can be reduced to\foot{In
the Euclidean case the poles from the density of state
are identified as the existence of the discrete states \HPT.
Even in our case there is contribution from the poles,
but it leads to the imaginary part contrary to the Euclidean case.}
\eqn\torusn{\eqalign{Z(q)&={\rm Tr}_{\cal H}
  q^{L_0 -\frac{c}{24}} \bar q^{\bar L_0 - \frac{c}{24}} \cr
  &= i V_{D-2} \int \frac{d^{D-2} p}{(2 \pi)^{D-2}}
   \int^{\infty}_{-\infty} d\ti{s} \sum_{m,\bar m,N,\bar N}
     \rho(\ti{s}) (q \bar q)^{\frac{\ti{s}^2}{\alpha}
    + \frac{(\vec p)^2}{4} - \frac{D-2}{24}}
     q^{\frac{m^2}{(2-\alpha)} + N}
     \bar q^{\frac{\bar m^2}{(2-\alpha)}
     + \bar N},
}}
where $N, \bar N$ represent physical closed string oscillators
and the volume of $D-2$ dimensional space is denoted as $V_{D-2}$.
The density of state $\rho(\ti{s})$ is defined by
\eqn\doss{\rho(\ti{s})= \frac{1}{2\pi}2\ln L
 - \frac{1}{2\pi } \frac{d}{2d\ti{s}}
 \ln \frac{\Gamma (\ha + \ti{s} -m) \Gamma (\ha +\ti{s} + \bar m)}
           {\Gamma (\ha - \ti{s} -m) \Gamma (\ha -\ti{s} + \bar m)} ,}
where we take $L \to \infty$ limit\foot{%
The similar formula can be also obtained for the usual Liouville
theory (say, at the value $\ap=2$ discussed later). In the
case of  $c\leq 1$ string theory (e.g. see a review \KlR),
the matrix model dual
reproduces only the bulk term (the first term in \doss)
since we can only see the quantities which depend on the cutoff $L$
due to the double scaling limit.}. This makes the first term
diverge, which reflects the infinite volume in the time direction.
The second term appears due to the time-dependence of the
background. We can see that the possible divergence arises from
the IR region $\tau_2 \to \infty$ (due to the tachyon)
and the poles of $\rho(\tilde s)$.

\subsec{IR Divergence due to Tachyonic Modes}

{}From the expression of \torusn , we can see that there is a divergence
from the IR region $\tau_2 \to \infty$ due to the existence of the
tachyonic modes $N=\bar N=0$.
Following  \Marcus\  we regularize the divergence by an
analytical continuation.
We first split the amplitude into the finite and divergent pieces as
\eqn\decayrate{Z = \int_{\cal F} \frac{d \tau_1 d\tau_2}{\tau_2} Z(q)
  = \int_{< R} \frac{d \tau_1 d\tau_2}{\tau_2} Z(q)
  + \int^{\frac12}_{-\frac12} d \tau_1 \int_{R}^{\infty}
   \frac{d\tau_2}{\tau_2} Z(q)}
with a cut off $R$.
Utilizing a formula \Marcus \CKL
\eqn\analytic{{\rm Im}\, \int^{\infty}_{R}
 \frac{d x}{x} x^{-n} e^{(a+i \epsilon)x}
 = \cases{\pi a^n /n! &\hbox{for $a \geq 0$},\cr
   0&\hbox{for $a<0$},}}
we find that the imaginary part comes from the second term as
\eqn\imaginary{{\rm Im}\, \frac{Z}{i V_{D-2}}
 = \pi \sum_{m} \int \frac{d^{D-2} p}{(2 \pi)^{D-2}}
  \int_0^{\hat s} ds \rho (\tilde s).
}
We have used
\eqn\hats{ \hat s  = \sqrt{ \alpha \left(\frac{D-2}{24} -
 \frac{m^2}{(2-\alpha)} - \frac{\vec p^2}{4}\right)},}
and the level matching condition $L_0 = \bar L_0$ coming from the
$\tau_1$ integration. The inside of the square root must be positive.
This imaginary part \imaginary\ implies the instability of the vacuum
due to the existence of the tachyon, which was discussed in \Marcus\ for
the flat spacetime and in \CKR\ for the time-like linear dilaton.
The novelty here is the modification of the density of the tachyon modes
due to the time-dependence.

\subsec{Emission of Closed Strings with High Energy}

Apart from the IR divergence discussed in the previous subsection,
there is another type of divergence coming from
the poles of the density of states \doss .
Since the structure of poles is complicated,
we consider only a special case $\ap=2$, where the gauged WZW model
is equivalent to the time-like Liouville theory and a free boson.
Practically we set $m=\bar m = 0$ and add another momentum $p_1$.
Then we obtain
\eqn\vacpoles{{\rm Im}\, \frac{Z}{i V_{D-2}}
 = 2\pi \s{2-\ap}\int_{\cal F} \frac{d \tau_1 d\tau_2}{\tau_2}
   \int \frac{d^{D-1} p}{(2\pi)^{D-1}}
 e^{-\pi\tau_2 \vec{p}^2}|\eta(\tau)|^{-2(D-2)}
\sum_{n=0}^{\infty} e^{-2\pi\tau_2(n+\f{1}{2})^2},} where we pick
up the poles located at $\tilde s = -1/2 -n$. We have used the
contour of integral of $\tilde s$ through just below the poles for
$\tilde s < 0$, just above for $\tilde s > 0$ and large $|\tilde
s|$ in the upper half plain, as is naturally derived from the
deformation of the counter \intt. The quantized
momentum $n$ in the time direction (see \vacpoles) may suggest a
periodicity along the imaginary time or a thermal property
as is similar to the decaying D-branes \LLM.

What is the physical interpretation of this vacuum instability?
This model has only closed strings as degrees of freedom, so
the energy coming from the vacuum decay has to be carried out
by closed strings.
Let us see the UV behavior of the imaginary part.
For the purpose, it is convenient to use the domain which is obtained
by $S$-transformation ($\tau \to -1/\tau$) from the
fundamental domain ${\cal F}$.
Employing the Poisson resummation formula
\eqn\pos{
\sum_{n=0}^{\infty}e^{-2\pi\tau_2(n+\f{1}{2})^2} =\f{1}{\s{\tau_2}}
\sum_{w=0}^{\infty}\rho_w e^{-\f{\pi w^2}{2\tau_2}}
 = \sum_{w=0}^{\infty} \rho_w
 \int^{\infty}_{-\infty} dp_0 e^{-\pi \tau_2 p_0^2
 + \sqrt2 \pi i p_0 w}}
with
\eqn\rhow{\rho_w =
 \cases{\frac{1}{2\sqrt2} &for $w = 0$, \cr
 \frac{(-1)^w}{\sqrt2} & otherwise,}}
we can see that in the UV region $\tau_2 \to 0$ the leading contribution
comes from $w=0$ sector.
Since the equation \vacpoles\ with \pos\ for $w=0$ becomes
the vacuum amplitude for the theory with $D$ free bosons and ghosts,
the amplitude diverges due to the
 Hagedorn tower of closed string states $\rho(\tau)\sim
e^{\f{\pi}{6}(D-2)\f{1}{\tau_2}}$ in the UV region.
Roughly speaking, this behavior seems consistent with the relation
\hagec\ between the two point function
$|\langle e^{-i\omega t}e^{-i\omega t}\rangle|=1$ and
the closed string emission rate.

\newsec{Possible Relation to Rolling Closed String Tachyon Condensation}

Up to now we have investigated on the one side of the FZZ duality,
namely, the $SL(2,\blackb{R})/U(1)$ WZW model with $0<k<2$. The
model itself, of course, deserves to be investigated as it describes
an interesting time-dependent background. The other side of the
FZZ duality is given by the sine-Liouville theory
\eqn\timel{S=\int dz^2\left[ \de X\bar{\de} X-\de
X^0\bar{\de}X^0+\lambda e^{\beta
X^0}\cos\left(\f{X}{R}\right)\right],} parameterized by
\eqn\params{Q=\f{i}{\s{\ap}},\ \ R=\f{1}{\s{2-\ap}},\ \
\beta=\s{\ap}.} This is the model of $1+1$ dimensional spacetime
(plus the $D-2$ dimensional flat space)
with the time-dependent closed string tachyon field and the linear
dilaton \eqn\clt{T(x_0,x)=\lambda e^{\s{\ap}X^0} \cos(\s{2-\ap}\
X), \qquad g_s=e^{-\f{X^0}{\s{\ap}}}.} In other words, the model
represents the system with the closed string tachyon condensing
inhomogeneously as time evolves. Therefore, we have a new solvable
model of the dynamical process of closed string tachyon
condensation via this duality. The closed string pair creation
driven by the closed string tachyon condensation \clt\ should be
dual to that in the time-dependent background \timedmd .

Because the FZZ duality is a kind of T-duality, we expect that the
geometry driven by the tachyon condensation is that of the T-dual
of the $SL(2,\blackb{R})/U(1)$ model \timedmd . The
cosmological background starts from the strongly coupled region, then
expands for a while, and finally approaches to a flat spacetime with
a small coupling constant. In the FZZ dual model, one starts in
the strongly coupled region with a small tachyon field, and ends up
with a weak coupling constant but with a large tachyon field. Then
one may wonder why the tachyon becomes so large in the late time,
though the FZZ-dual metric \timedmd\ approaches to the flat
spacetime.

We can propose an answer\foot{A similar issue appears in the two
dimensional string theory (or $c=1$ string) by regarding the $b\to
1/b$ duality \ZZL\ in the ordinary Liouville theory as a sort of
FZZ duality in our sine-Liouville case. The time-dependence of the
cosmological constant $\sim e^{2x^0}$ is opposite to that of the
dual one $\sim e^{-2x^0}$. In the recent work \TTL\ done after
this paper, some evidences  that there is actually no such paradox
were obtained by using the $c=1$ matrix model. In that context the
dual cosmological constant is not relevant for physical
quantities.} to
 this question by considering the choice of the vacuum.
In such a time-dependent background the choice of the vacuum is very
important in general. As is clear from the minisuperspace
analysis, we have infinitely many kinds of vacua by assigning
various boundary conditions for the propagating field. In our
case, we chose a specific vacuum with the smooth boundary
condition at $t=0$ in the $SL(2,\blackb{R})/U(1)$ model. We
regarded this vacuum as the correct one because the choice of the
vacuum leads to the exact two point functions \twol . In other
words, our simple Lorentzian continuation selects this vacuum
among other many possible ones. In the sine-Liouville theory side, this
corresponds to a specific choice of 'vacuum'\foot{At the
singularity it is difficult to define the positive (or negative)
frequency modes. We regarded the boundary condition as the
definition of the vacuum. This vacuum can be formally written as
$|vac\lb _{t=0} \propto \prod_{\omega}
e^{-\frac12 S(\omega)a^{\dagger}_\omega a^{\dagger}_\omega}|0\lb _{t=\infty}$
\SO \GS \ST\
and in our case
its normalization is singular since the two point function $S(\omega)$
is a pure phase.} at the singularity $(t=0)$,
and the tachyon field might have a large value at the late time.
Indeed, the two point function \twol\ has poles which can be
explained only by the tachyon field \clt\ regarded as a screening
operator. If we use the wave function $\Psi\equiv
g_s^{-1}\Phi$ that is
canonical in the (sine-)Liouville theory \Seiberg, then the boundary
condition can be rephrased into the requirement $\Psi(t=0)=0$. This
seems to suggest that the real singularity at $t=0$ can not be
seen from fundamental strings, which might give an interesting resolution
of the cosmological singularity in string theory.

We should also note that we are considering the region $0<k<2$
where the quantum effects are important.
In fact, even in the Euclidean $SL(2,\blackb{R})/U(1)$ model,
the quantum effects may change the classical picture.
The wave function \Seiberg\ of the
sine-Liouville perturbation $\Phi\sim e^{(Q-1/Q)\phi}$
becomes large in the weakly coupled
region for small $k\,(>2)$, which is opposite to the classical case
with large $k$. We believe that this fact is also closely related to the
behavior of the tachyon field discussed just before.

\newsec{Conclusions and Discussions}

In this paper we investigated the $SL(2,\blackb{R})/U(1)$ WZW
model with level $0<k<2$ as an exactly solvable time-dependent
background in bosonic string theory. We examined the exact
geometry of the model and analyzed the minisuperspace model.
We read the two point functions from the difference of the two
vacua (at the starting point and the far future) in the minisuperspace
computations and
found a match to the exact two point functions in the conformal
field theory. In other words, the exact two point function given
by continuing the Euclidean model ($k>2$) to the Lorentzian one
($0<k<2$) gives an important information on the choice of vacua.
The non-zero two point function implies the closed string pair
production, and we estimated the total production rate. We found
that it diverges exponentially, which means that there is a very
large backreaction at one-loop level. Indeed we showed that the
imaginary part of its one-loop amplitude includes a similar
divergence. We also calculated the exact three point function.

In addition to the advantage that various exact quantities can be
obtained in the conformal field theory level, we have the exact
metric including stringy corrections. This allows us to examine
the cosmological structure of the two dimensional time-dependent
spacetime. Furthermore, the model is expected to be T-dual (FZZ
dual) to a background with an inhomogeneous and time-dependent
closed string tachyon. Indeed we showed that the two point function
in time-like bulk Liouville theory can be correctly
reproduced as a particular
case of our model. This duality suggests that the model with
the rolling closed string tachyon is equivalent to the string
theory on the cosmological background. This may lead to an
interesting geometrical understanding of closed string tachyon
condensation.

Another cosmological model can be defined by the Lorentzian
$SL(2,\blackb{R})/U(1)$ WZW model with a negative level $k<0$.
This model has also Lorentzian time-dependent
geometry as a target space as investigated in \TV \KL \LU \CKR \BLW.
 Applying the duality  to the WZW
model, we may obtain the model with the following closed string
tachyon field \eqn\dualt{T=e^{\s{|k-2|}X^0}
\cosh\left(\sqrt{|k|} X \right).} The continuation of the original
results for $k>2$ to this case $k<0$ seems to be highly
non-trivial since
the $X$ direction becomes decompactified in the negative level
case. It would be interesting to investigate also this model
furthermore.

\vskip 0.4in

\centerline{\bf Acknowledgments}

The authors are grateful to S.~Minwalla,
S.~Mukohyama, S.-J.~Rey, A.~Strominger and H.~Takayanagi for useful
discussions. YH would like to thank the members of KEK for their
hospitality and useful discussions. The work of TT was supported
in part by DOE grant DE-FG02-91ER40654.

\listrefs

\end